\documentstyle[graphicx,prl,aps,amsmath]{revtex} \textwidth 15cm

\newcommand{\doublint}{\int\rule{-3.5mm}{0mm}\int} 
 
\newcommand{\vecbm}[1]{\mbox{\boldmath$#1$}}

\newcommand{\lra} {$\leftrightarrow$}
\newcommand{\vecb}[1]{\mbox{\bf#1}} 
\newcommand{\lora}{{\boldmath$\longrightarrow$}}
\begin{document}
\paperwidth =15cm
%%\twocolumn
% \draft command makes pacs numbers print
\draft
%\tighten
\title{Straight way to Thermo-Statistics,
Phase Transitions, Second Law of
Thermodynamics, but without
Thermodynamic Limit} 
\author{D.H.E. Gross} \address{Hahn-Meitner-Institut
  Berlin, Bereich Theoretische Physik,Glienickerstr.100\\ 14109
  Berlin, Germany and Freie Universit{\"a}t Berlin, Fachbereich
  Physik; \today} 
\maketitle
\begin{abstract}
  Boltzmann's principle $S(E,N,V)=k\ln W$ relates the entropy to the
  geometric area $e^{S(E,N,V)}$ of the manifold of constant energy in
  the (finite-$N$)-body phase space. From the principle, all
thermodynamics and especially all phenomena of phase transitions and
critical phenomena can be deduced. The topology of the curvature
matrix $C(E,N)$ (Hessian) of $S(E,N)$ determines regions of pure
phases, regions of phase separation, and (multi-)critical points and
lines. Thus, $C(E,N)$ describes all kind of phase-transitions with
all their flavor. They are linked to convex (upwards bending)
intruders of $S(E,N)$, here the canonical ensemble defined by the
Laplace transform to the intensive variables becomes multi-modal,
non-local, and violates the basic conservation laws (it mixes widely
different conserved quantities). The one-to-one mapping of the
Legendre transform gets lost.  Within Boltzmann's principle,
Statistical Mechanics becomes a {\em geometric theory} addressing the
whole ensemble or the manifold of {\em all} points in phase space
which are consistent with the few macroscopic conserved control
parameters. Moreover, this interpretation leads to a straight
derivation of irreversibility and the Second Law of Thermodynamics
out of the time-reversible microscopic mechanical dynamics. It is the
whole ensemble that spreads irreversibly over the accessible phase
space not the single $N$-body trajectory.  This is all possible
without invoking the thermodynamic limit, extensivity, or concavity
of $S(E,N,V)$ and also without invoking any cosmological constraints.
Without the thermodynamic limit or at phase-transitions the systems
are usually not self-averaging, do not have a single peaked
distribution in phase space.  It is further shown that non-extensive
Hamiltonian systems at equilibrium are described by Boltzmann's
principle and {\em not} by Tsallis non-extensive statistics.
\end{abstract}
\section{Introduction}
There are many attempts to derive Statistical Mechanics from first
principles. The earliest are by
Boltzmann~\cite{boltzmann,boltzmann1877,boltzmann1884,boltzmann23a},
Gibbs~\cite{gibbs02,gibbs36}, and
Einstein~\cite{einstein02,einstein03,einstein04,einstein05a}. The two
central issues of Statistical Mechanics according to the deep and
illuminating article by Lebowitz~\cite{lebowitz99a} are to explain
how irreversibility (the Second Law of Thermodynamics) arises from
fully reversible microscopic dynamics, and the other astonishing
phenomenon of Statistical Mechanics: the occurence of phase
transitions. In this paper I want to present an easy, straightforward
derivation of both aspects directly out of the microscopic
time-reversal invariant Newton-mechanics invoking a minimum of
assumptions. We will see how both problems are connected.

There is an important aspect of Statistical Mechanics which to my
opinion was not sufficiently considered up to now: Statistical
Mechanics and also Thermodynamics are {\em macroscopic} theories
describing the {\em average}~\footnote{Here I do not speak of the {\em
    typical} behavior. This would only be the same if the system is
  {\em self-averaging}, which I do not demand, see below.}  behavior
of {\em all} $N$-body systems with the {\em same macroscopic
  constraints}.  It is this fact and nothing else that leads in a
simple and straightforward manner to the desired understanding of
irreversibility, the Second Law for $N$-body systems, which obey a
completely time reversible Hamiltonian dynamics, and leads
simultaneously to the full spectrum of phase-transition phenomena.  It
is certainly essential to deduce the Second Law from reversible (here
Newtonian) and not from dissipative dynamics as is often done because
just the {\em derivation of irreversibility} from fully {\em
  reversible} dynamics is the main mystery of Statistical Mechanics.
Here a first hint: Whereas a single trajectory in the (finite-$N$)-body
phase space returns after a {\em finite} Poincar\'{e} recurrence time
a manifold of points develops in general irreversibly with time, see
below.

\section{Minimum-bias deduction of Statistical Mechanics}
Thermodynamics presents an economic but reduced description of a
$N$-body system with a typical size of $N\sim 10^{23}$ particles in
terms of a very few ($M\sim 3-8$) ``macroscopic'' degrees of freedom
($dof$'s). Here we will allow also for much smaller systems of some
$100$ particles like nucleons in a nucleus. However, I assume that
always $6N\gg M$.  The believe that phase transitions and the Second
Law can exist only in the thermodynamic limit will turn out to be
false.

Evidently, determining only $M$ $dof$'s leaves the overwhelming number
$6N-M$ $dof$'s undetermined. {\em All} N-body systems with the same
macroscopic constraints are {\em simultaneously} described by
Thermodynamics. These systems define an {\em ensemble} $\cal{M}$ of
points~\footnote{In this paper I denote ensembles or manifold in phase
  space by calligraphic letters like ${\cal{M}}$.} in the $N$-body
phase space. Thermodynamics can only describe the {\em average}
behavior of this whole group of systems. I.e.\ it is a {\em
  statistical} or {\em probabilistic} theory. Considered on this level
we call Thermodynamics thermo-statistics or since Gibbs {\em
  Statistical Mechanics}.

The dynamics of the (eventually interacting) $N$-body system is ruled
by its Hamiltonian $\hat{H}_N$. Let us in the following assume that
our system is trapped in an inert rectangular box of volume $V$ and
there is no further conservation law than the total energy. The
motion in time of all points of the ensemble follows trajectories in
$N$-body phase space $\{q_i(t),p_i(t)\}|_{i=1}^N$ (I consider only
classical mechanics) which will never leave the ($6N-1$)-dimensional
shell (or manifold) $\cal{E}$ of constant energy $E$ in phase space.
We call this manifold the {\em micro-canonical} ensemble.

An important information which contains the whole equilibrium
Statistical Mechanics including all phase transition phenomena is the
area $W(E,N) =:e^S$ of this manifold $\cal{E}$ in the $n$-body phase
space.  Boltzmann has shown that $S(E,N,V)$ is the {\em entropy} of our
system.  Thus the entropy and with it equilibrium thermodynamics has
a {\em geometric} interpretation.

Einstein called Boltzmann's definition of entropy as e.g.\ written on
his famous epitaph
\begin{equation}
\fbox{\fbox{\vecbm{$S=k$\cdot$lnW$}}}\label{boltzmentr1}\end{equation}
{\em Boltzmann's principle}~\cite{einstein05d} from which Boltzmann was
able to deduce thermodynamics. Precisely $W$ is the number of
micro-states~\footnote{In the following I will call single points in
the $6N$-dim phase-space {\em states} or micro-states which are
specific microscopic realizations of the $N$-body system and
correspond to single N-body quantum states in quantum mechanics.
These must be distinguished from {\em macro-states} used in
phenomenological thermodynamics c.f.\ section~\ref{EPSformulation}.}
of the $N$-body system at given energy $E$ in the spatial volume $V$
and further-on I put Boltzmann's constant $k=1$:
\begin{eqnarray}
W(E,N,V)&=& tr[\epsilon_0\delta(E-\hat H_N)]\label{partitsum}\\
tr[\delta(E-\hat H_N)]&=&\int_{\{q\in V\}}{\frac{1}{N!}
\left(\frac{d^3q\;d^3p}
{(2\pi\hbar)^3}\right)^N\delta(E-\hat H_N)},\label{phasespintegr}
\end{eqnarray} 
$\epsilon_0$ is a suitable energy constant to make $W$ dimensionless,
the $N$ positions $q$ are restricted to the volume $V$, whereas the
momenta $p$ are unrestricted.  In what follows, I remain on the level
of classical mechanics. The only reminders of the underlying quantum
mechanics are the measure of the phase space in units of $2\pi\hbar$
and the factor $1/N!$ which respects the indistinguishability of the
particles (Gibbs paradoxon). With this definition,
eq.(\ref{boltzmentr1}), {\em the entropy $S(E,N,V)$ is an everywhere
multiple differentiable, one-valued function of its arguments.} This
is certainly not the least important difference to the conventional
canonical definition.

In contrast to Boltzmann~\cite{boltzmann1877,boltzmann1884} who used
the principle only for dilute gases and to
Schr\"odinger~\cite{schroedinger44}, who thought equation
(\ref{boltzmentr1}) is useless otherwise, I take the principle as {\em
  the fundamental, generic definition of entropy}. In a recent
book~\cite{gross174} cf.\ also~\cite{gross173,gross175} I demonstrated
that this definition of thermo-statistics works well especially also
at higher densities and at phase transitions {\em without invoking the
  thermodynamic limit}. This is important: Elliot
Lieb~\cite{lieb97,lieb98a} considers the additivity of $S(E)$ and
Lebowitz~\cite{lebowitz99,lebowitz99a} the thermodynamic limit as
essential for the deduction of thermo-statistics. However, neither is
demanded if one starts from Boltzmann's principle. {\em Boltzmann's
  principle eq.(\ref{boltzmentr1}) is the only axiomatic assumption
  necessary for thermo-statistics.}

{\em This is all that Statistical Mechanics demands}, no further
assumption must be invoked. Neither does one need extensivity
\footnote{Dividing extensive systems into larger pieces, the total
  energy and entropy are equal to the sum of those of the pieces. I
will call non-extensive systems where this is not the case in the
following also ``Small'' systems~\cite{gross174} (with a capital
$S$!) to stress the paradoxical point that the largest systems
in nature (globular galaxies) belong to this
group as well, nevertheless, they cannot be treated in the
thermodynamic limit.}, nor additivity, nor concavity of $S(E)$
c.f.~\cite{lavanda90}. In the next section I will show how
Boltzmann's principle allows to define phase-transitions in
``Small'', non-extensive as well in normal ``large'' extensive
systems where our more general definition of phase transitions (see
below) will coincide with the conventional definition by the Yang-Lee
singularities~\cite{yang52,huang64}. 

Of course one should not wonder if some familiar features of
conventional canonical thermo-statistics do not hold anymore in
``Small'' systems. This is discussed in some more detail in
subsection~\ref{equivalence} 
\section{Why is the micro-canonical ensemble fundamental?} \label{why} 
During the dynamical evolution of a many-body system interacting by
short range forces the internal energy is conserved.  Only
perturbations by an external ``container'' can change the energy.
I.e.\ the fluctuations of the energy are
\begin{equation}
\frac{\Delta E}{E}\propto V^{-1/3}.
\end{equation}  
If, however, the diameter of the system is of the order of the range
of the force, i.e.\ the system is ``Small'', non-extensive, details of
the coupling to the container cannot be ignored. 

\subsection{Equivalence of ensembles and self-averaging
\label{equivalence}} 

In contrast, the canonical ensemble does not care about these
details, assumes the system is homogeneous, averages over a
Boltzmann-Gibbs (exponential) distribution
$P_{BG}\{q_\alpha,p_\alpha\}=\frac{1}{Z(\beta)}e^{-\beta \hat
H\{q_\alpha,p_\alpha\}}$ of energy and fixes only the mean value of
the energy by the temperature $1/\beta$. In order to agree with the
micro, $e^{-\beta E}W(E)$ must be sharp in $E$ i.e.  self-averaging,
what is usually not the case in non-extensive systems or at phase
transitions of first order.  Then one {\em must} work in the micro,
the only orthode ensemble.  The micro-ensemble assumes precise --
perhaps idealized -- boundary conditions for each particle
independently of whether the system is small or large. Therefore,
already Gibbs considered the micro-ensemble as the fundamental and
the canonical as approximation to it. He demonstrates clearly the
failure of the canonical in cases of phase separation or other
situations where both ensemble differ, footnote on page 75
of~\cite{gibbs02}, see also~\cite{ehrenfest12,ehrenfest12a}.

There are important features where the microcanonical statistics of
``Small'' systems deviates from the ``canonical'' structure of
conventional thermo-statistics of extensive systems in the
thermodynamical limit:

E.g. {\em the familiar Legendre-transform structure, a paradigm of
  ``canonical'' thermo-statistics, is lost}.  Clearly, without
self-averaging, fixing an intensive parameter like the temperature $T$
does not fix the energy sharply. Most evident example is a transition
of first order in the canonical ensemble at the transition temperature
where the energy per particle fluctuates by the specific latent heat
even in the thermodynamic limit. Related is the occurrence of negative
specific heat cf. section~\ref{wrongcurv} found in recent experiments
on nuclei~\cite{gross171,dAgostino00} which was predicted many years
before~\cite{gross95}. Here, there are at least three energies for the
same temperature c.f. section~\ref{systphas}. The present discussions
of non-extensive statistics as proposed by Tsallis~\cite{tsallis88} or
recently by Vives et al.~\cite{vives01} clearly miss this crucial
point. In the Tsallis statistics the entropy is expressed by the
mean-values of the extensive quantities like
$<\!\!E\!\!>$~\cite{vives01,martinez00a} controlled by a Lagrange
parameter $\beta$ or $\beta^*$. Of course, this is equivalent to the
micro-ensemble (is an orthode) only if the variance of the energy is
small.  In one or the other way the thermodynamic limit, extensivity,
and self-averaging is still demanded where Legendre transformations
(may) become one to one.  However, in the case of non-extensive systems
the existence of the thermodynamic limit is unlikely and so is the
uniqueness of the Legendre transformation.

\subsection{Tsallis statistics does not apply to Hamiltonian systems
at equilibrium} 

Tsallis suggested to extend the Boltzmann-Gibbs {\em canonical} energy
distribution $P_{BG}\{q_i,p_i\}$ by using the
$q$-exponential~\cite{abe01,martinez00a}:
\begin{eqnarray}
e_q^x&=&[1+(1-q)x]^{\frac{1}{1-q}}\label{e_q}\\
\mbox{and replacing }P_{BG}\{q_i,p_i\}&=&\frac{e^{-\beta
H\{q_i,p_i\}}}{Z(\beta)}\\
\mbox{where: }Z(\beta)&=&\int_{\{q_i\in V\}}{\frac{1}{N!}
\left(\frac{d^3q_i\;d^3p_i}
{(2\pi\hbar)^3}\right)^Ne^{-\beta H\{q_i,p_i\}}}\\
\mbox{with mean-values }<\!\!O\!\!>_{BG}&=&\int_{\{q_i\in V\}}{\frac{1}{N!}
\left(\frac{d^3q_i\;d^3p_i}
{(2\pi\hbar)^3}\right)^NO(q_i,p_i)\;P_{BG}(q_i,p_i)}\\
\mbox{by: }P_{BG}\{q_i,p_i\}&\to&[f\{q_i,p_i\}]^q\nonumber\\
\mbox{where: }f\{q_i,p_i\}&=&\frac{e_q^{-\beta H\{q_i,p_i\}}}{Z_q(\beta)}
\label{p_q}\\
\mbox{and }Z_q(\beta)&=&\int_{\{q_i\in V\}}{\frac{1}{N!}
\left(\frac{d^3q_i\;d^3p_i}
{(2\pi\hbar)^3}\right)^N\;e_q^{-\beta H\{q_i,p_i\}}}\\
\mbox{and calculating mean values }<\!\!O\!\!>_q&=&\frac{N_q}{D_q}\\
N_q&=&\int_{\{q_i\in V\}}{\frac{1}{N!}
\left(\frac{d^3q_i\;d^3p_i}
{(2\pi\hbar)^3}\right)^NO(q_i,p_i)[f(q_i,p_i)]^q},\\
D_q&=&\int_{\{q_i\in V\}}{\frac{1}{N!}
\left(\frac{d^3q_i\;d^3p_i}
{(2\pi\hbar)^3}\right)^N[f(q_i,p_i)]^q},\\
\lim_{q\to1}[f\{q_i,p_i\}]^q&=&\frac{e^{-\beta H\{q_i,p_i\}}}{Z(\beta)},
\end{eqnarray}
which leads to the Tsallis
$q$-entropy~\cite{tsallis88,martinez00a,martinez00,abe01}:
\begin{equation}
S_q=k\frac{1-\sum_{\alpha=1}^W{P_\alpha^q}}{q-1}\hspace{1cm}
\lim_{q\to 1}S_q=-\sum{P_\alpha\ln{P_\alpha}}\label{q-entropy}.
\end{equation}
The Tsallis $q$-entropy is very similar to $q$-dimension well known in
mathematics~\cite{crc99}. Its main purpose is to
emphasize/suppress small probabilities $P^q_\alpha$ depending on the
parameter $q<1$ or $q>1$ resp..

For a closed {\em Hamiltonian} system at energy $E$, the $P_\alpha$
are the probabilities for each of the $W(E)$ microscopic
configurations (quantum states). According to Toral~\cite{toral01} this
has of course the following consequences: After maximizing $S_q(E)$
under variation of $P_\alpha$ with the constraint of
$\sum{P_\alpha}=1$ one obtains the equal probability distribution
characterized by Boltzmann's entropy
$W(E)=e^{S(E)}$:\begin{equation}P_\alpha=\left\{\begin{array}{ll}
      e^{-S(E)}&,\epsilon_\alpha=E\\ 0&,otherwise
    \end{array}\right.,\hspace{1cm}S_q=
  k\frac{1-e^{(1-q)S(E)}}{q-1}.\end{equation}
Moreover, following Abe~\cite{abe01} and Toral~\cite{toral01} the original
definitions of the microcanonical temperature and pressure~(\ref{TP})
through Boltzmann's entropy $S(E,N,V)$, eq.(\ref{boltzmentr1}), are
the only way {\em within} Tsallis statistics to define the equilibrium of
two systems in weak contact and to fulfill the Zeroth Law under
energy- and volume exchange see 
also~\cite{vives01}:\begin{eqnarray}S(E,N,V)&=&\ln W(E,N,V)\\
  T_{phys}=\left(\frac{\partial S}{\partial
      E}\right)^{-1}&\hspace{1cm}&P_{phys}=\frac{\partial S/\partial
    V}{\partial S/\partial E}\label{TP}.\end{eqnarray}I.e. the physical
quantity {\em relevant for equilibrium} of {\em Hamiltonian} systems,
extensive or not, is the original Boltzmann entropy $S(E)=\ln[W(E)]$,
eq.(\ref{boltzmentr1}), whatever the non-extensivity index $q$.
 Therefore, for closed {\em Hamiltonian}
many-body systems {\em at statistical equilibrium}, extensive or
not,{\em the thermo-statistical behavior is entirely controlled by
  Boltzmann's principle and the microcanonical ensemble as discussed
  in this paper.} Tsallis statistics seems to apply to non-equilibrium 
situations like turbulence~\cite{beck01} or the border of 
chaos~\cite{latora99} etc.
\section{Phase transitions within Boltzmann's principle}

At phase-separation the system becomes inhomogeneous and splits into
different regions with different structure. This is the main generic
effect of phase transitions of first order.  Evidently, phase
transitions are foreign to the (grand-) canonical theory with
homogeneous density distributions.  In the conventional Yang-Lee
theory phase transitions~\cite{yang52} are indicated by the zeros of
the grand-canonical partition sum where the grand-canonical formalism
breaks down because of the Yang--Lee singularities of the
grand-canonical potentials. In the following I show that the
micro-canonical ensemble gives a much more detailed and more natural
insight which moreover corresponds to the experimental identification
of phase transitions by interfaces (inhomogeneities).
\subsection{Relation of the topology of $S(E,N,V)$ to the Yang-Lee 
zeros of $Z(T,\mu,V$)} The grand-canonical partition sum may be
obtained out of the micro-canonical one by a double Laplace
transform. To explore the link to the Yang-Lee singularities I
discuss it for the moment in the thermodynamic limit (large volume V,
large number of particles $N$ but homogeneous constant density
$n=N/V$. In this limit it does not matter whether $N$ is discrete or
continuous.)

\begin{eqnarray}
Z(T,\mu,V)&=&\doublint_0^{\infty}{\frac{dE}{\epsilon_0}\;dN\;e^{-[E-\mu
N-TS(E)]/T}}\nonumber\\
&=&\frac{V^2}{\epsilon_0}\doublint_0^{\infty}{de\;dn\;e^{-V[e-\mu n-T
s(e,n)]/T}}\label{grandsum}\\
&\approx&\frac{V^2}{\epsilon_0}\doublint_0^{\infty}de\;dn\;
e^{-V[\mbox{\scriptsize const.+lin.+quadr.}]}\nonumber
\end{eqnarray}
and we investigate the specific free energy $f(T,\mu)=-[\ln(Z)]/V$ in
the thermodynamic limit $V\to\infty|_{N/V= const}$.

The double Laplace integral (\ref{grandsum}) can be evaluated
asymptotically for large $V$ by expanding the exponent as indicated
in the third line to second order in $\Delta e,\Delta n$ around the
``stationary point'' $e_s,n_s$ where the linear terms vanish:
\begin{eqnarray}
\frac{1}{T}&=&\left.\frac{\partial s}{\partial e}\right|_s\nonumber\\
\frac{\mu}{T}&=&-\left.\frac{\partial s}{\partial n}\right|_s
\label{statpoint}
\end{eqnarray} 
the only terms remaining to be integrated are the quadratic ones.
{\em If the eigen-curvature $\lambda_1<0$ eq.(\ref{curvdet}), and
  eqs.(\ref{statpoint}) have a single solution ($e_s,n_s$)}, this is
then a Gaussian integral and yields:
\begin{eqnarray}
Z(T,\mu,V)&=&\frac{V^2}{\epsilon_0}e^{-V[e_s-\mu
n_s-T{\mbox{\boldmath$\scriptstyle s(e_s,n_s)$}}]/T}
\doublint_{-\infty}^{\infty}{dv_1\;dv_2\;
e^{V[\lambda_1 {\mbox{\boldmath$\scriptstyle
v$}}_1^2+\lambda_2{\mbox{\boldmath$\scriptstyle v$}}_2^2]/2}}\\
&=&e^{-F(T,\mu,V)/T}\\ f(T,\mu,V):=\frac{F(T,\mu,V)}{V}&\to& e_s-\mu
n_s-Ts_s
+\frac{T\ln{(\sqrt{\det(e_s,n_s)})}}{V}+o(\frac{\ln{V}}{V})\label{asympt}.
\end{eqnarray}
Here $\det(e_s,n_s)$ is the determinant of the {\em curvatures}
(Hessian) of $s(e,n)$, $\vecbm{v_1,v_2}$ are the eigenvectors of
the Hessian.

\begin{equation}
\det(e,n)= \left\|\begin{array}{cc}
\frac{\partial^2 s}{\partial e^2}& \frac{\partial^2 s}{\partial n\partial e}\\
\frac{\partial^2 s}{\partial e\partial n}& \frac{\partial^2 s}{\partial n^2}
\end{array}\right\|= \left\|\begin{array}{cc} s_{ee}&s_{en}\\
s_{ne}&s_{nn}
\end{array}\right\|=\lambda_1\lambda_2,\hspace{1cm}\lambda_1\ge\lambda_2
 \label{curvdet}
\end{equation}
$\lambda_1$ can be positive or negative. If $\lambda_1<0$ and
eqs.(\ref{statpoint}) have no other solution, the last two terms in
eq.(\ref{asympt}) go to $0$, and we obtain in the thermodynamic limit
($V\to\infty$) the familiar result for the free energy density:
\begin{equation}
f(T,\mu,V\to\infty)=e_s-\mu n_s-Ts_s.
\end{equation} 
I.e.\ the {\em curvature $\lambda_1$ of the entropy surface $s(e,n,V)$
  or the largest eigenvalue of the Hessian matrix decides whether the
  grand-canonical ensemble agrees with the fundamental micro-ensemble
  in the thermodynamic limit.} If this is the case and
eqs.(\ref{statpoint}) have a single solution or $s(e,n)$ touches its
concave hull at $e_s,n_s$, then there is a pointwise one to one
mapping of the micro-canonical entropy $s(e,n)$ to the grand-canonical
partition sum $Z(T,\mu)$, and $\ln[Z(T,\mu)]/V$ or $f(T,\mu)$ is
analytical in $z=e^{\beta\mu}$. Due to Yang and Lee we have then a
single, stable phase~\cite{huang64}. Otherwise, {\em the Yang-Lee
  zeros of $Z(T,\mu)$ reflect anomalous points/regions of
  $\lambda_1\ge 0$} ($\det(e,n)\le 0$) where the {\em canonical
  partition sum does not reflect local properties of the
  micro-ensemble, i.e.\ does not respect the conservation laws, and
  mixes conserved quantities}.  This is crucial.  As $\det(e_s,n_s)$
can be studied for finite or even small systems as well, this is the
only proper extension of phase transitions to ``Small'' systems.

\subsection{The physical origin of the wrong curvature\label{wrongcurv}}

I will now discuss the physical origin of the convex (upwards
bending) intruders in the entropy surface for systems with
short-range coupling in two examples.
\subsubsection{Liquid-gas transition in sodium clusters \label{Nacluster}}
\begin{figure}
\begin{minipage}[b]{6cm}
\includegraphics*[bb = 99 57 400 286, angle=-0, width=6cm,  
clip=true]{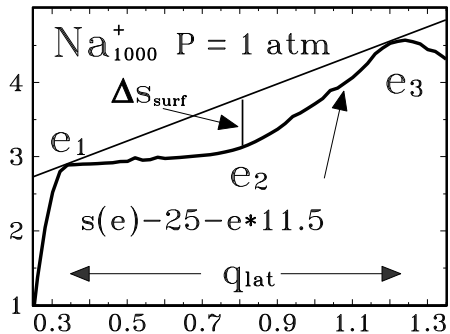}%%%}
\caption{MMMC~\protect\cite{gross174} simulation of the entropy 
  $s(e)$ per atom ($e$ in eV per atom) of a system of $N_0=1000$
  sodium atoms with an external pressure of 1 atm.  At the energy
  $e_1$ the system is in the pure liquid phase and at $e_3$ in the
  pure gas phase, of\label{naprl0f}}
\end{minipage}\hspace*{0.5cm}\begin{minipage}[b]{8cm}
\small{course with fluctuations. The latent heat per atom is
$q_{lat}=e_3-e_1$.  Attention: the curve $s(e)$ is artifically
sheared by subtracting a linear function $25+e*11.5$ in order to make
the convex intruder visible. {\em $s(e)$ is always a steeply
monotonic rising function}.We clearly see the global concave
(downwards bending) nature of $s(e)$ and its convex intruder. Its
depth is the entropy loss due to the additional correlations by the
interfaces. It scales $\propto N^{-1/3}$. From this one can calculate
the surface tension per surface atom $\sigma_{surf}/T_{tr}=\Delta
s_{surf}*N_0/N_{surf}$.  The double tangent (Gibbs construction) is
the concave hull of $s(e)$. Its derivative gives the Maxwell line in
the caloric curve $T(e)$ at $T_{tr}$. In the thermodynamic limit the
intruder would disappear and $s(e)$ would approach the double tangent
from below. Even though, however, the probability of configurations
with phase-separations are suppressed by the (infinitely small)
factor $e^{-N^{2/3}}$ relative to the pure phases and the
distribution remains {\em strictly bimodal in the canonical
ensemble}.}
\end{minipage}
\end{figure}

In table (\ref{tab}) I compare the ``liquid--gas'' phase transition
in sodium clusters of a few hundred atoms with that of the bulk at 1
atm.\ c.f.\ also fig.(\ref{naprl0f}).\\
\hspace*{0.5cm}\begin{minipage}[h]{6cm}
\begin{table}
\begin{tabular} {|c|c|c|c|c|c|} \hline 
&$N_0$&$200$&$1000$&$3000$&\vecb{bulk}\\ %\tableline
\hline 
\hline  
&$T_{tr} \;[K]$&$940$&$990$&$1095$&\vecb{1156}\\ \cline{2-6}
&$q_{lat} \;[eV]$&$0.82$&$0.91$&$0.94$&\vecb{0.923}\\ \cline{2-6}
{\bf Na}&$s_{boil}$&$10.1$&$10.7$&$9.9$&\vecb{9.267}\\ \cline{2-6}
&$\Delta s_{surf}$&$0.55$&$0.56$&$0.44$&\\ \cline{2-6}
&$N_{surf}$&$39.94$&$98.53$&$186.6$&$\vecbm{\infty}$\\ \cline{2-6}
&$\sigma/T_{tr}$&$2.75$&$5.68$&$7.07$&\vecb{7.41}\\
\hline
\end{tabular}
\end{table}
\end{minipage}\hspace*{0.5cm}\begin{minipage}[h]{8cm}
\begin{table}
\caption{Parameters of the liquid--gas transition of small sodium 
clusters (MMMC-calculation) in comparison with the bulk for rising 
number $N_0$ of atoms, $N_{surf}$ is the average number of surface 
atoms (estimated here as $\sum{N_{cluster}^{2/3}}$) of all clusters 
together. $\sigma/T_{tr}=\Delta s_{surf}*N_0/N_{surf}$ corresponds 
to the surface tension. Its bulk value is adjusted to the 
experimental input values used for the binding energies of clusters 
as given by Brechignac et al. c.f.\protect\cite{gross174}.
\label{tab}} \end{table}
\end{minipage}
Conclusion: For systems with short range interactions a convex
intruder in $s(e,n)$ appears with the fragmentation of the system
into several clusters and monomers. The number of surface particles
scales with the depth of the intruder (surface entropy). I.e.\ the
convex intruder signals the preference of the system to become
inhomogeneous, the characteristic  signal for the separation of different
phases (liquid and gas) at a phase transition of first order.
\subsubsection{The topology of the entropy surface $S(E,N)$ for Potts
lattice gases} 

Having discussed in the previous example a system with a single
thermodynamic degree of freedom or control parameter (the energy $E$)
we will now study more subtle features.  If the system has two, or
more, degrees of freedom, e.g. energy $E=Ve$ and particle number
$N=Vn$, where $V$ is the volume, we can have phase boundaries and
critical points. This is similar to the classical $P-V$ diagram of the
liquid--gas phase transition in the Van der Waals theory. We are now
able to identify multi-critical points. These were previously studied
in the canonical ensemble only, where sophisticated finite size
scaling is needed to identify these points.

As example we investigate the 3-states diluted Potts model on a {\em
finite} 2-dim (here $L^2=50^2$) lattice with periodic boundaries in
order to minimize effects of the external surfaces of the system. The
model is defined by the Hamiltonian:
\begin{eqnarray}
H&=&-\sum_{i,j}^{n.n.pairs}o_i o_j\delta_{\sigma_i,\sigma_j}\\
n&=&L^{-2}N=L^{-2}\sum_io_i .\nonumber
\end{eqnarray}
Each lattice site $i$ is either occupied by a particle with spin
$\sigma_i =-1,0,\mbox{ or }1$, or it is empty (vacancy).  The sum is
over pairs of neighboring lattice sites $i,j$, and the occupation
numbers are:
\begin{equation}
o_i=\left\{\begin{array}{cl}
1&\mbox{, spin particle in site }i\\
0&\mbox{, vacancy in site }i\\
\end{array}\right. .
\end{equation}

This model is an extension of the ordinary ($q=3$)-Potts model to
allow also for vacancies. At zero concentration of vacancies ($n=1$),
the system has in the thermodynamic limit a continuous phase
transition at $e_c=1+\frac{1}{\sqrt{q}}\approx 1.58$
\cite{baxter73,pathria72}. With rising number of vacancies the
probability to find a pair of particles at neighboring sites with the
same spin orientation decreases. I.e.  this is similar to a larger
number of spin orientation $q_{eff}$ on each lattice site in the
ordinary Potts model, where we know that the transition of second
order becomes a transition of first order for $q>4$. Thus the
inclusion of vacancies has the effect of an increasing effective
$q_{eff}\ge 3$.  This results in an increase of the critical energy of
the continuous phase transition with decreasing $n$ and provides a
line of continuous transition, which is supposed to terminate when
$q_{eff}$ becomes larger than $4$. From here on the transition becomes
first order.

At smaller energies the system is in one of three ordered phases
(spins predominantly parallel in one of the three possible
directions).

\begin{figure}\begin{center}
\includegraphics*[bb =0 0 290 180, angle=-0, width=9cm,  
clip=true]{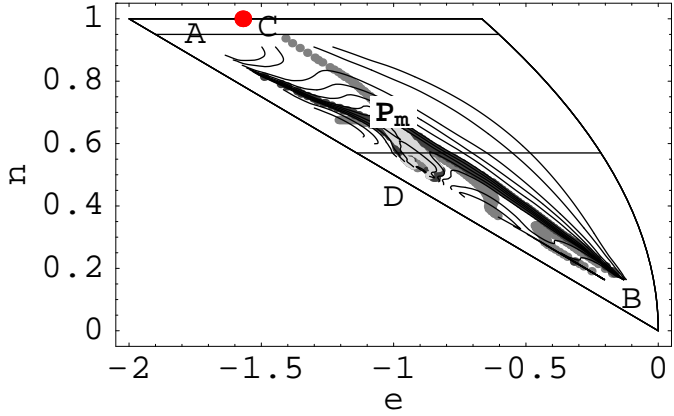}%{potconts.eps}
%{potcontdens.eps}%{P3detgrdetcont.eps}%{p3detg3.eps}%{P3detgrdetcont.eps}
\end{center}
\caption{Conture plot of the curvature determinant, eq.(\ref{curvdet}), 
  of the 2-dim Potts-3 lattice gas with $50*50$ lattice points, $n$ is
  the number of particles per lattice point, $e$ is the total energy
  per lattice point; Dark grey lines: $\det=0$, boundary of the region
  of phase coexistence ($\det<0$), in the triangle $AP_mB$; Light grey
  line: minimum of $\det(e,n)$ in the direction of the largest
  curvature ($\vecb{v}_{\lambda_{max}}\cdot\vecbm{\nabla}\det=0$),
  line of second order transition; In the triangle $AP_mC$ pure
  ordered (solid) phase ($\det>0$); Above and right of the line
  $CP_mB$ pure disordered (gas) phase ($\det>0$); The crossing $P_m$
  of the boundary lines is a multi-critical point. It is also the
  critical end-point of the region of phase separation ($\det<0$). The
  light gray region around the multi-critical point $P_m$ corresponds
  to a flat region of $\det(e,n)\sim 0$ i.e.
  {\boldmath$\vecbm{\nabla}$} \mbox{$\lambda_1$}{\boldmath$\sim 0$},
  details see \protect\cite{gross173}; $C$ is the critical point of
  the ordinary Potts, $q=3$ model (without vacancies) {\em in the
    thermodynamic limit}.} \label{det}
\end{figure}

Figure (\ref{det}) shows how for a small system of $50*50$ lattice
points all phenomena of phase transitions can be studied from the
topology of the determinant of curvatures (\ref{curvdet}) in the
micro-canonical ensemble. In this example the second curvature
$\lambda_2<0$ so $\mbox{sign}(\det)=-\mbox{sign}(\lambda_1)$.
\subsection{Systematics of phase transitions in the
micro-ensemble~\label{systphas}} Now we can give a systematic and
generic classification of phase transitions in terms of the topology
of curvatures of $s(e,n)$ which applies also to ``Small'' systems:
\begin{itemize}
\item \begin{minipage}[t]{8.5cm} A {\bf single stable} phase by
    $\lambda_1<0$. Here $s(e,n)$ is locally concave (downwards bended)
    in both directions and eqs.(\ref{statpoint}) have a single solution
    $e_s,n_s$. Then there is a one to one mapping of the
    grand-canonical \lra the micro-ensemble. The order parameter is
    the direction $\vecb{v}_1$ of the eigenvector of largest curvature
    $\lambda_1$. In many situations one may have only locally
    $\lambda_1<0$, however, there may be further solutions to
    eqs.(\ref{statpoint}) farther away. In such cases which have no
    equivalent in the canonical ensemble, {\em we will still speak of
      regions in \{$e,n$\} of pure phases embedded in regions of
      phase-separation.}
\end{minipage}~~~\begin{minipage}[t]{7cm} %\vspace{-0.5cm}
%!%\includegraphics*[bb = 117 13 490 645, angle=-90, width=5.5cm,
  \includegraphics*[bb = 61 6 429 624, angle=-90, width=5.5cm,
  clip=true]{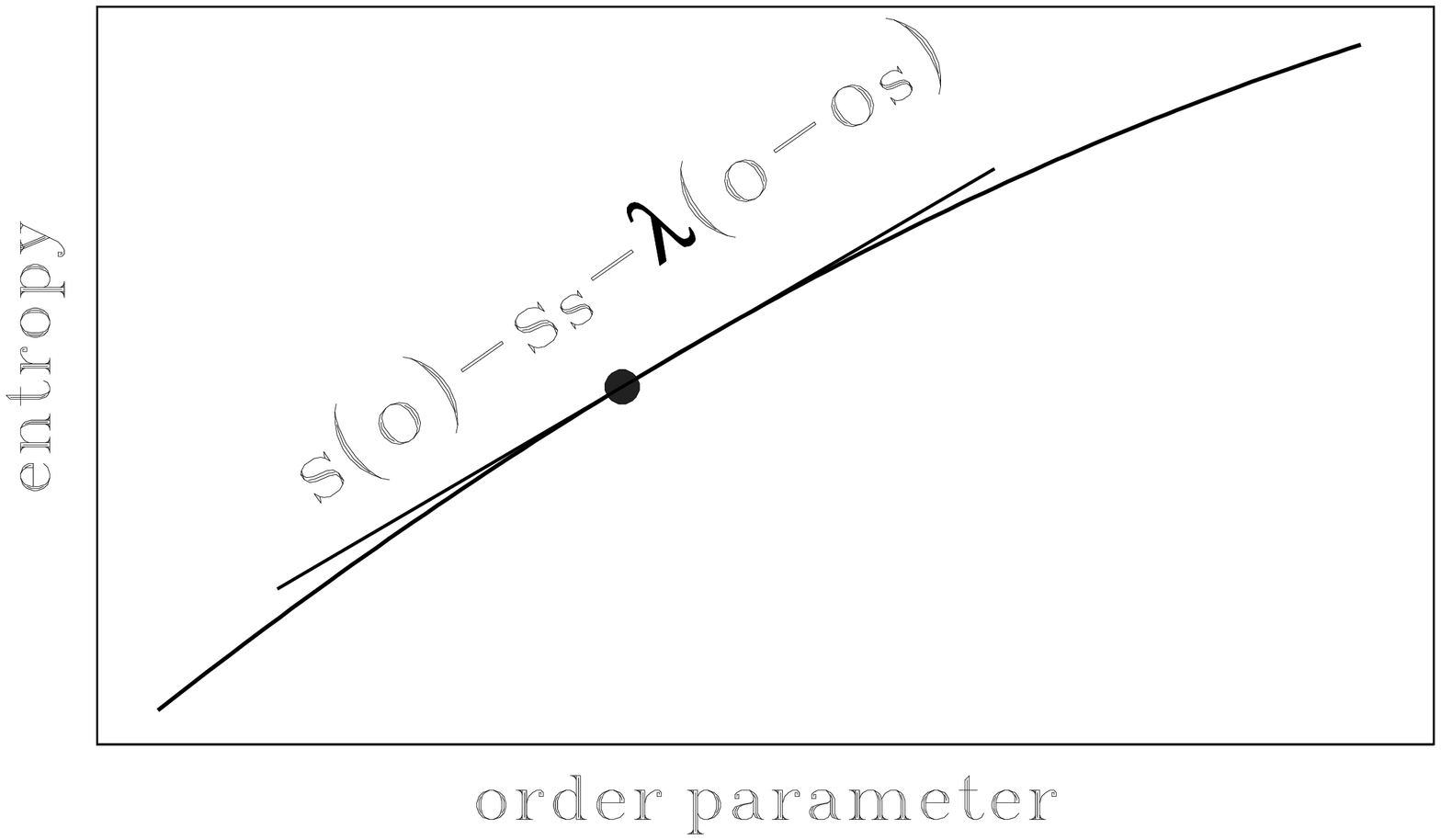}\end{minipage}
\item \begin{minipage}[t]{8.5cm} A {\bf transition of first order}
    with phase separation and surface tension (c.f.subsection
    \ref{Nacluster}) indicated by $\lambda_1>0$. $s(e,n)$ has a
convex intruder (upwards bended) in the direction $\vecb{v}_1$ of the
largest curvature.  Then eqs.(\ref{statpoint}) have multiple
solutions, at least three.  The whole convex area of \{e,n\} is
mapped into a single point ($T,\mu$) in the grand-canonical ensemble
(non-locality)\label{convex}. I.e.\ if the largest curvature of
$S(E,N)$ is $\lambda_1>0$ {\em both ensembles are not equivalent, the
(grand-) canonical ensemble is non-local in the order parameter and
violates basic}
\end{minipage}~~~\begin{minipage}[t]{7cm} %!%\vspace{-0.5cm}
%!%\includegraphics*[bb = 130 8 490 640, angle=-90, width=6cm
  \includegraphics*[bb = 91 12 462 630, angle=-90, width=5.5cm,
  clip=true]{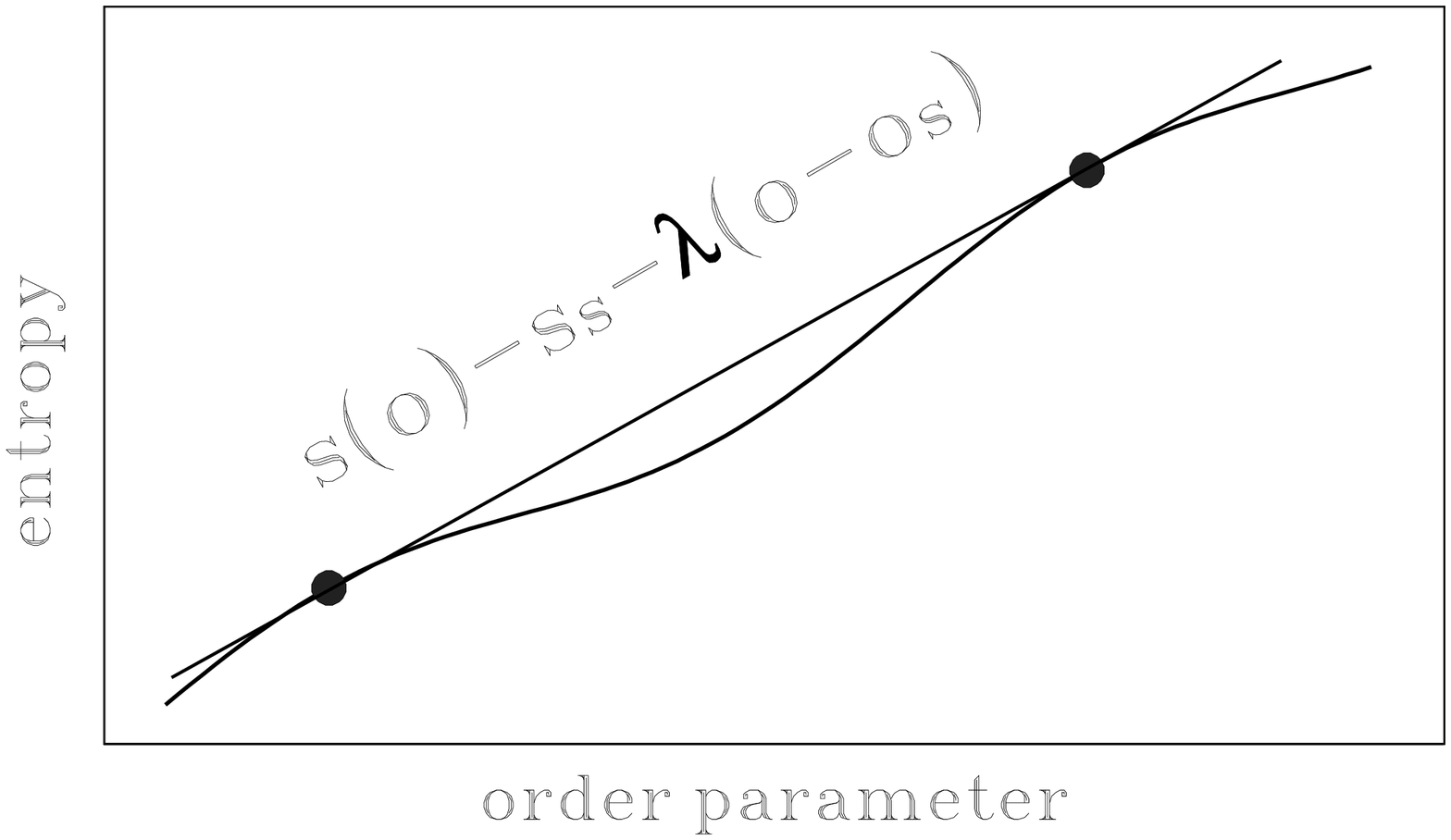} \end{minipage}\\{\em conservation laws.}  See
also~\cite{gross174,gross173,gross175,gross176}. The region in the
plane of conserved control-parameters $e,n$ where we have a separation
of different phases, $\lambda_1(e,n)>0$, is bounded by lines with
$\lambda_1(e,n)=0$. On this boundary is the end-point of the
transition of first order,
\item where we have a {\bf continuous (``second order'')} transition
  with vanishing surface tension, where two neighboring phases become
  indistinguishable. This is at points where the two stationary points
  move into one another to become the critical end-point of the first
  order transition. This is then also a maximum of $\lambda_1$. I.e.\
where $\lambda_1(e,n)=0$ and
$\vecb{v}_{\lambda_1=0}\cdot\vecbm{\nabla}\lambda_1=0$.  These are
the {\em catastrophes} of the Laplace transform $E\to T$. Here
$\vecb{v}_{\lambda_1=0}$ is the eigenvector of the Hessian belonging
to the largest curvature eigenvalue $\lambda_1=0$.  ($\vecb{v}_1$
plays the role of the order parameter of the transition. In this
direction one moves fastest from one phase to the other.)  At the
other points of $\lambda_1(e,n)=0$ one of the two coexisting phases
gets depleted.  Furthermore, there may be also whole lines of
second-order transitions like in the anti-ferro-magnetic Ising model
c.f.\cite{gross174}.
\item Finally, a {\bf multi-critical point} where more than two phases
  become indistinguishable is at the branching of several lines in the
  \{$e,n$\}-phase-diagram with
  $\lambda_1=0$,{\boldmath$\vecbm{\nabla}$}
  \mbox{$\lambda_1$}{\boldmath$= 0$}.  Fig.~\ref{det} gives an
  illustration of a multicritical point in a small system.
\end{itemize}

Just a comment on the interesting suggestion by Pettini et
al.~\cite{casetti01} to characterize phase transitions by topological
changes of the potential energy: They claim that phase transitions
are signalized by sudden changes of the topology of the
configurational energy.

However, one has to keep in mind: In the thermodynamic limit, first
order transitions are characterized by two essential things:
\begin{enumerate}
\item There are two or more phases (different types of configurations,
  liquid and gas, or ordered and disordered) which {\em coexist at
the same temperature} $T=[\partial S/\partial E]^{-1}$.
\item Between both configurations the total energy per particle, or
the potential energy per particle differs by the specific latent
heat.
\end{enumerate}
Of course not every topological jump in the configuration energy
corresponds to a phase transition~\cite{casetti01} and not every jump
is accompanied by the same kinetic energy per particle (temperature)
or more precisely by the same slope $T=[\partial S/\partial E]^{-1}$
of the entropy. Evidently, topological changes are likely {\em
  neccessary} conditions (Pettini) for phase transitions to occur,
however, the {\em sufficient} condition is the same temperature
$T=[\partial S/\partial E]^{-1}$ at two different values of the
order-parameters, energy or magnetization etc.

Also the suggestion of Borrmann et al.~\cite{muelken01} to
characterize phase transitions in small systems by the zeros of the
canonical partition sum must be commented: This idea extends the idea
of Yang and Lee and the work of Grossmann~\cite{grossmann69}.  First,
the canonical partition sum is a {\em derived} quantity which is not
an orthode at phase transitions, c.f.\ section~\ref{why}, as was
already noticed by Gibbs, footnote on page 75 of~\cite{gibbs02}.
Moreover, in some situations of non-extensive systems it may not even
exist.
\section{Measuring a macroscopic observable}
\label{EPSformulation}

After succeeding to deduce equilibrium statistics including all
phenomena of phase transitions from Boltzmann's principle alone even
for ``Small'' systems, i.e.\ non-extensive many-body
systems~\cite{gross174}, it is challenging to explore how far this
``most conservative and restrictive way to
thermodynamics''~\cite{bricmont00} is able to describe also the {\em
approach} of (possible ``Small'') systems to equilibrium and the
Second Law of Thermodynamics.

Thermodynamics describes the development of {\em macroscopic}
features of many-body systems without specifying them microscopically
in all details. Before I address the Second Law, I have to clarify
what I mean with the label ``macroscopic observable''.

A single point $\{q_i(t),p_i(t)\}_{i=1,\cdots,N}$ in the $N$-body
phase space corresponds to a detailed specification of the system
with all degrees of freedom ($dof$'s) completely fixed at time $t$
(microscopic determination, the curly brackets indicate the whole set
of $6N$ coordinates $q_i,p_i$ of all particles $i$).  Fixing only the
total energy $E$ of an $N$-body system leaves the other
($6N-1$)-degrees of freedom unspecified.  A second system with the
same energy is most likely not in the same microscopic state as the
first, it will be at another point in phase space, the other $dof$'s
will be different.  I.e.\ the measurement of the total energy
$\hat{H}_N$, or any other macroscopic observable $\hat{M}$,
determines a ($6N-1$)-dimensional {\em sub-manifold} ${\cal{E}}$ or
${\cal{M}}$ in phase space. (The manifold ${\cal{M}}$ is called by
Lebowitz a {\em macro-state}~\cite{lebowitz99,lebowitz99a} which
contains $\Gamma_M=W(M)$ micro-states. I, however, prefer to use the
name ``state'' only for micro-states or points in phase space.)  All
points (the micro-states) in $N$-body phase space consistent with the
given value of $E$ and volume $V$, i.e.\ all points in the
($6N-1$)-dimensional sub-manifold ${\cal{E}}(N,V)$ of phase space are
equally consistent with this measurement. ${\cal{E}}(N,V)$ is the
micro-canonical ensemble. This example tells us that {\em any
macroscopic measurement is incomplete and defines a sub-manifold of
points in phase space not a single point}. An additional measurement
of another macroscopic quantity $\hat{B}\{q,p\}$ reduces ${\cal{E}}$
further to the cross-section ${\cal{E}\cap\cal{B}}$, a
($6N-2$)-dimensional subset of points in ${\cal{E}}$ with the volume:
\begin{equation}
W(B,E,N,V)=\frac{1}{N!}\int{\left(\frac{d^3q\;d^3p}
{(2\pi\hbar)^3}\right)^N\epsilon_0\delta(E-\hat H_N\{q,p\})\;
\delta(B-\hat B\{q,p\})}
\label{integrM}\end{equation}
If $\hat H_N\{q,p\}$ as well as also $\hat B\{q,p\}$ are continuous
differentiable functions of their arguments, what I assume in the
following, ${\cal{E}}\cap{\cal{B}}$ is closed. In the following I use
$W$ for the Riemann or Liouville volume (Hausdorff measure) of a
many-fold.

Microcanonical thermo{\em statics} gives the probability $P(B,E,N,V)$
to find the $N$-body system in the sub-manifold
${\cal{E}\cap\cal{B}}(E,N,V)$:
\begin{equation}
P(B,E,N,V)=\frac{W(B,E,N,V)}{W(E,N,V)}=e^{\ln[W(B,E,N,V)]-S(E,N,V)}
\label{EPS}\end{equation}
This is what Krylov seems to have had in mind~\cite{krylov79} and
what I will call the ``ensemble probabilistic formulation of
Statistical Mechanics ($EPS$) ''.

Similarly thermo{\em dynamics} describes the development of some
macroscopic observable $\hat{B}\{q_t,p_t\}$ in time of systems which
were specified at an earlier time $t_0$ by another macroscopic
measurement $\hat{A}\{q_0,p_0\}$.  It is related to the volume of the
sub-manifold
${\cal{M}}(t,t_0)={\cal{A}}(t_0)\cap{\cal{B}}(t)\cap{\cal{E}}$:
\begin{equation} W(A,B,E,t)=\frac{1}{N!}\int{\left(\frac{d^3q_t\;d^3p_t}
{(2\pi\hbar)^3}\right)^N\delta(B-\hat B\{q_t,p_t\})\;
\delta(A-\hat A\{q_0,p_0\})\;\epsilon_0\;\delta(E-\hat H\{q_t,p_t\})},
\label{wab}
\end{equation}
where $\{q_t\{q_0,p_0\},p_t\{q_0,p_0\}\}$ is the set of trajectories
solving the Hamilton-Jacobi equations
\begin{equation}
\dot{q}_i=\frac{\partial\hat H}{\partial p_i},\hspace{1cm}
\dot{p}_i=-\frac{\partial\hat H}{\partial q_i},\hspace{1cm}i=1\cdots N
\end{equation}
with the initial conditions $\{q(t=t_0)=q_0;\;p(t=t_0)=p_0\}$.  For a
large system with $N\sim 10^{23}$ the probability to find a given
value $B(t)$, $P(B(t))$, is usually sharply peaked as function of $B$
at its typical value. Such systems are called self-averaging.
Ordinary thermodynamics treats systems in the thermodynamic limit
$N\to\infty$ and gives only $<\!\!B(t)\!\!>$. However, here we are
interested to formulate the Second Law for ``Small'' systems i.e.\ we
are interested in the whole distribution $P(B(t))$ not only in its
mean value $<\!\!B(t)\!\!>$. There are also many situation where the
system is not self-averaging, where a finite variance remains even in
the thermodynamic limit. (E.g.\ at phase transitions of first order
the energy per particle fluctuates in the canonical ensemble by the
specific latent heat.)

There is an important property of macroscopic measurements: Whereas at
finite times Hamilton dynamics evolves a compact region of phase space
again into a compact region, this does not need to be so at infinite
times.  Then, at $t\to\infty$, the set may not be closed anymore
(perhaps a fractal, see below). This means there exist series of
points $\{a_n\}\in{\cal{A}}(t=\infty)$ which converge to a point
$\lim_{n\to\infty}a_n=:a_{n=\infty}$ which is {\em not} in
${\cal{A}}(t=\infty)$. E.g.\ such points
$a_{n=\infty}\notin{\cal{A}}(\infty)$ may have intruded from the phase
space complementary to ${\cal{A}}(t_0)$.  Illustrative examples for
this evolution of an initially compact sub-manifold into a fractal set
are the generalized baker transformations discussed in this context by
ref.~\cite{fox98,gilbert00}. See reference~\cite{crc99} for the
fractal distribution produced by the general baker transformation.(As
any housewife knows, a baker dough becomes an infinitely thin
(fractal) puff pastry after pounding and folding it infinitely often.)
Then no macroscopic (incomplete) measurement can resolve
$a_{n=\infty}\notin{\cal{A}}(t=\infty)$ from its immediate neighbors
$a_n\in{\cal{A}}(t=\infty)$ in phase space with distances
$|a_n-a_{n=\infty}|$ less then any arbitrary small $\delta$.  In other
words, {\em at the time $t-t_0\to\infty$ no macroscopic measurement
  with its incomplete information about
  $\{q_{t=\infty},p_{t=\infty}\}$ can decide whether
  $\{q_0\{q_{t=\infty},p_{t=\infty}\},p_0\{q_{t=\infty},
  p_{t=\infty}\}\}\in{\cal{A}}(t_0)$ or not.} I.e.\ any macroscopic
theory like thermodynamics can only deal with the {\em closure} of
${\cal{A}}({t\to\infty})$. If necessary, the sub-manifold
${\cal{A}}({t\to\infty})$ must be artificially closed~\footnote{First
  $t\to\infty$ then the closure, not the other way round c.f.\ 
  however, the discussion in~\ref{conclus}.} to
$\overline{{\cal{A}}({t=\infty})}$ as developed further in
section~\ref{fractSL}.  {\em Clearly, in this approach this is the
  physical origin of irreversibility.}

\section{On Einstein's objections against the EPS-probability} 
\label{einsteinEPS}
Before I proceed I must comment on Einstein's attitude to the
principle~\cite{pais82}: Originally, Boltzmann called $W$ the
``Wahrscheinlichkeit'' (probability), i.e.\ the relative time a
system spends (along a time-dependent path) in a given region of
$6N$-dim.\ phase space. Our interpretation of $W$ to be the number of
``complexions'' (Boltzmann's second interpretation) or quantum states
(trace) with the same energy was criticized by
Einstein~\cite{einstein05d} as artificial. It is exactly that
criticized interpretation of $W$ which I use here and which works so
excellently~\cite{gross174}.

According to Abraham Pais: ``Subtle is the Lord''~\cite{pais82},
Einstein was critical with regard to the definition of relative
probabilities by eq.(\ref{EPS}), Boltzmann's counting of
``complexions''. He considered it as artificial and not corresponding
to the immediate picture of probability used in the actual problem:
``The word probability is used in a sense that does not conform to
its definition as given in the theory of probability.  In particular,
cases of equal probability are often hypothetically defined in
instances where the theoretical pictures used are sufficiently
definite to give a deduction rather than a hypothetical
assertion''~\cite{einstein05d}.  He preferred to define probability
by the relative time a system (a trajectory of a single point moving
with time in the $N$-body phase space) spends in a given subset of
the phase space.  However, is this really the immediate picture of
probability used in Statistical Mechanics?  This definition demands
the ergodicity of the trajectory in phase space.  As I discussed
above, thermodynamics as any other macroscopic theory handles
incomplete, macroscopic informations of the $N$-body system. It
handles, consequently, the temporal evolution of {\em finite sized
sub-manifolds} - ensembles - not single points in phase space. In the
case of a very large system the {\em typical} outcomes of macroscopic
measurements are calculated.  Nobody waits in a macroscopic
measurement, e.g.\ of the temperature of a gas, long enough that an
atom can cross the whole system.

In this respect, I think the EPS version of Statistical Mechanics is
closer to the experimental situation than the duration-time of a
single trajectory. Moreover, in an experiment on a small system like
an excited nucleus, which then may fragment statistically later on,
the {\em average} over a multiple {\em repetition} of scattering
events is taken. No ergodic covering of the whole phase space by a
single trajectory in time is demanded. Fragmenting nuclei at such
high excitation  have a too short lifetime. This is analogous to the
statistics of a falling ball on a Galton's nail-board where also a
single trajectory does not touch all nails but is random. Only after
many repetitions the smooth binomial distribution is established. For
the discussion of the Second Law in {\em finite} systems, this is the
correct scenario, not the time average over a single ergodic
trajectory.

\section{Fractal distributions in phase space, Second Law}\label{fractSL}
Here I will first describe a simple working-scheme (i.e.\ a
sufficient method) which allows to deduce mathematically the Second
Law. Later, I will show how this method is necessarily implied by the
reduced information obtainable by macroscopic measurements or
theories.

Let us examine the following Gedanken experiment: Suppose the
probability to find our system at points $\{q_t,p_t\}_1^N$ in phase
space is uniformly distributed for times $t<t_0$ over the
sub-manifold ${\cal{E}}(N,V_1)$ of the $N$-body phase space at energy
$E$ and spatial volume $V_1$. At time $t>t_0$ we allow the system to
spread over the larger volume $V_2>V_1$ without changing its energy.
If the system is {\em dynamically mixing}, the majority of
trajectories $\{q_t,p_t\}_1^N$ in phase space starting from points
$\{q_0,p_0\}_1^N$ with $q_0\in V_1$ at $t_0$ will now spread over the
larger volume $V_2$.  Of course the Liouvillean measure of the
distribution ${\cal{M}}(t,t_0)$ in phase space at $t>t_0$ will remain
the same ($=tr[{\cal{E}}(N,V_1)]$)~\cite{goldstein59}. (The label
$\{q_0\in V_1\}$ of the integral means that the positions
$\{q_0\}_1^N$ are restricted to the volume $V_1$, whereas the momenta
$\{p_0\}_1^N$ are unrestricted.)
\begin{eqnarray}
\left.tr[{\cal{M}}(t,t_0)]
\right|_{\{q_0\in V_1\}}
&=&\int_{\{q_0\{q_t,p_t\}\in
V_1\}}{\frac{1}{N!}\left(\frac{d^3q_t\;d^3p_t}
{(2\pi\hbar)^3}\right)^N\epsilon_0\delta(E-\hat
H_N\{q_t,p_t\})}\nonumber\\ &=&\int_{\{q_0\in
V_1\}}{\frac{1}{N!}\left(\frac{d^3q_0\;
d^3p_0}{(2\pi\hbar)^3}\right)^N\epsilon_0\delta(E-\hat
H_N\{q_0,p_0\})},\\
\mbox{because of: }\frac{\partial\{q_t,p_t\}}{\partial\{q_0,p_0\}}&=&1.
\label{liouville}\end{eqnarray}
But as already argued by Gibbs the distribution ${\cal{M}}(t,t_0)$
will be filamented like ink in water and will approach any point of
${\cal{E}}(N,V_2)$ arbitrarily close.
$\lim_{t\to\infty}{\cal{M}}(t,t_0)$ becomes dense in the new, larger
${\cal{E}}(N,V_2)$.  The closure $\overline{{\cal{M}}(t=\infty,t_0)}$
becomes equal to ${\cal{E}}(N,V_2)$.  This is clearly expressed by
Lebowitz~\cite{lebowitz99,lebowitz99a}.
 
\begin{figure}
\begin{minipage}[t]{6cm}
\begin{center}$V_a$\hspace{2cm}$V_b$\end{center}
\vspace*{0.2cm}
\includegraphics*[bb = 0 0 404 404, angle=-0, width=5.7cm,
clip=true]{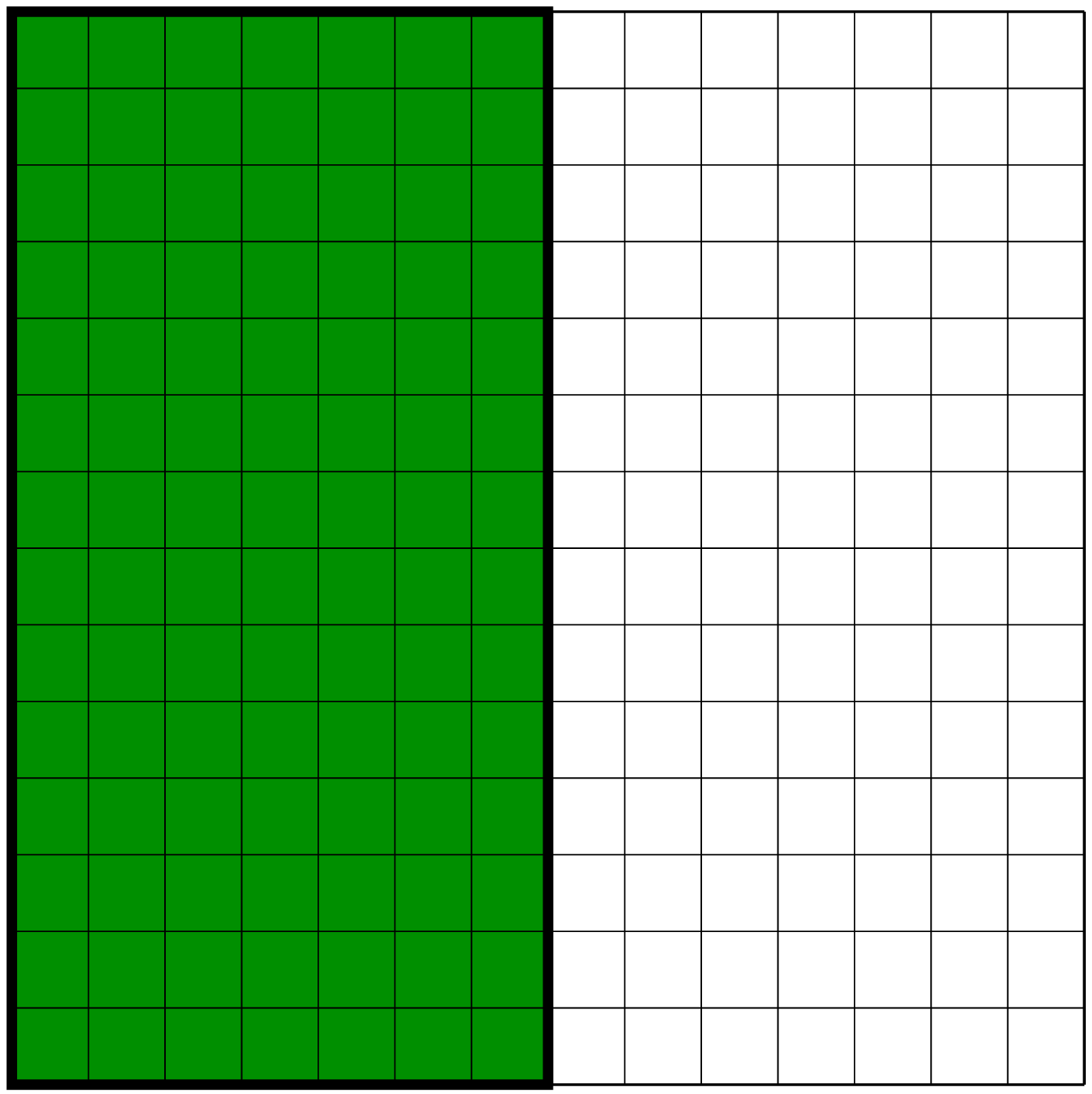}\begin{center}$t<t_0$\end{center}
\end{minipage}\lora\begin{minipage}[t]{6cm}
\begin{center}$V_a+V_b$\end{center}\vspace{-0.8cm}
\includegraphics*[bb = 0 0 490 481, angle=-0, width=6.9cm,
clip=true]{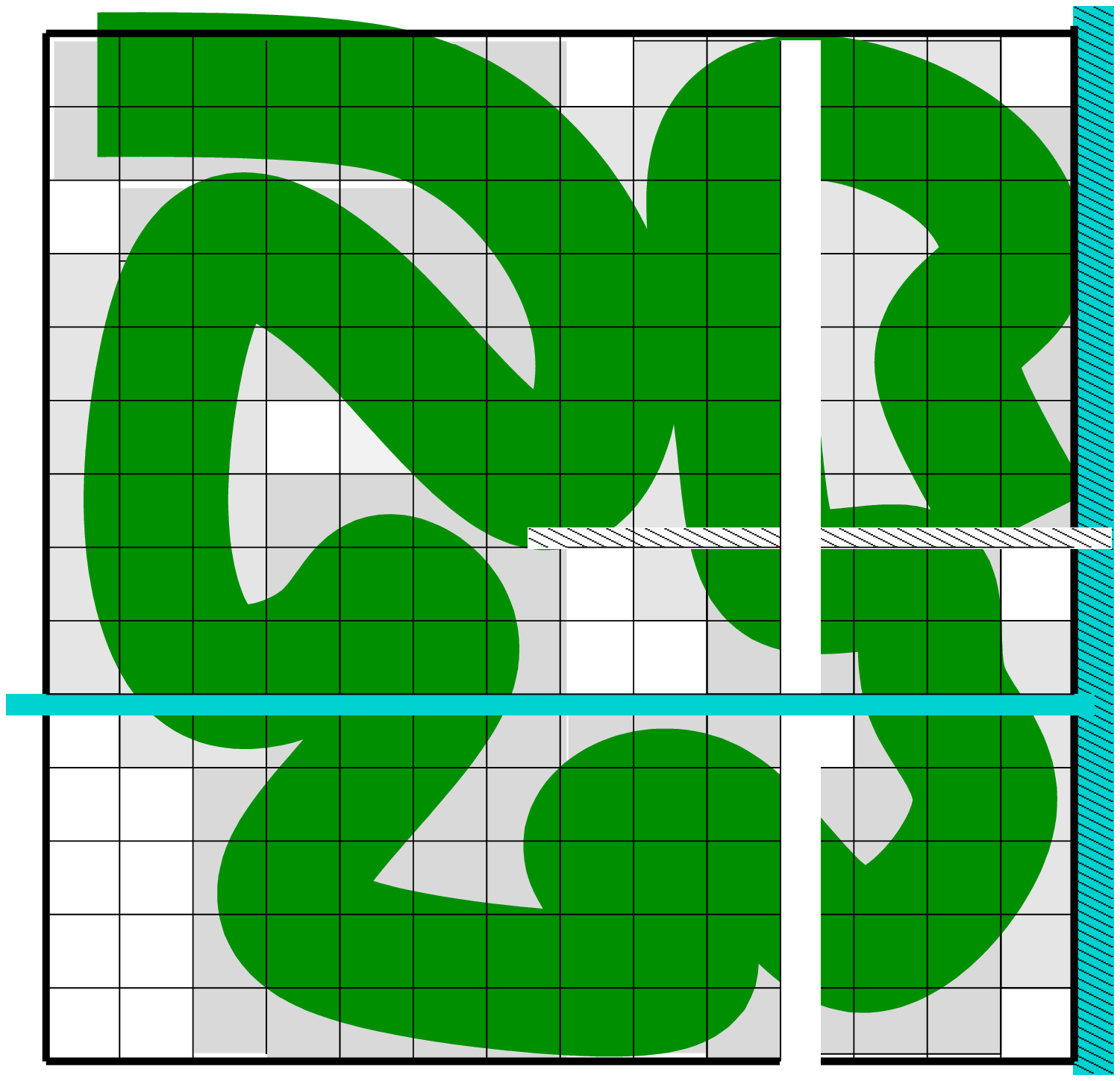}\begin{center}$t>t_0$\end{center}
\end{minipage}\\
\caption{The compact set ${\cal{M}}(t_0)$, left side, develops into an
  increasingly folded ``spaghetti''-like distribution
${\cal{M}}(t,t_0)$ in the phase-space with rising time $t$.  The
right figure shows only the early form of the distribution. At much
later times it will become more and more fractal and finally dense in
the new phase space.  The grid illustrates the boxes of the
box-counting method.  All boxes which overlap with ${\cal{M}}(t,t_0)$
contribute to the box-counting volume $\mbox{vol}_{box,\delta}$ and
are shaded gray.  Their number is $N_\delta$ \label{spaghetti}}
\end{figure}

In order to express this fact mathematically, I transform integrals
over the phase space like (\ref{phasespintegr}) or (\ref{integrM}):
\begin{equation}
W(E,N,t,t_0)=
\frac{1}{N!}\int_{\{q_0\{q_t,p_t\}\subset V_1\}}{\left(\frac{d^3q_t\;d^3p_t}
{(2\pi\hbar)^3}\right)^N\epsilon_0\delta(E-\hat H_N\{q_t,p_t\})}
\end{equation}
into:
\begin{eqnarray}
\int{\left(d^3q_t\;d^3p_t\right)^N\cdots}&=&
\int{d\sigma_1\cdots d\sigma_{6N}\cdots}\\
d\sigma_{6N}&:=&\frac{1}{||\nabla\hat H||}
\sum_i{\left(\frac{\partial\hat H}{\partial
q_i}dq_i+\frac{\partial\hat H}{\partial p_i}dp_i\right)}=
 \frac{1}{||\nabla\hat H||}dE\\
||\nabla\hat H||&=&\sqrt{\sum_i{\left(\frac{\partial\hat H}{\partial
q_i}\right)^2+\sum_i{\left(\frac{\partial\hat H}{\partial p_i}\right)^2}}}\\
W(E,N,t,t_0)&=&\frac{1}{N!(2\pi\hbar)^{3N}}
\int_{\{q_0\{q_t,p_t\}\subset V_1\}}
{d\sigma_1\cdots d\sigma_{6N-1}
\frac{\epsilon_0}{||\nabla\hat H||}}.
\end{eqnarray}
{\em Now, I redefine Boltzmann's definition of entropy
eq.(\ref{boltzmentr1} to \ref{phasespintegr})}:
\begin{eqnarray}
S&=&\ln(W(E,N,V)\\
W(E,N,V)&=& tr[\epsilon_0\delta(E-\hat H_N)]\\
tr[\delta(E-\hat H_N)]&=&\int_{\{q\in V\}}{\frac{1}{N!}
\left(\frac{d^3q\;d^3p}
{(2\pi\hbar)^3}\right)^N\delta(E-\hat H_N)},
\end{eqnarray} 
by replacing the Riemannian integral for $W$ by its box-counting
``measure'': 
\begin{equation}
W(E,N,V)\to \displaystyle{B_d\hspace{-0.5 cm}
\int}_{\{q_0\{q_t,p_t\}\in V_1\}}
{\frac{1}{N!}\left(\frac{d^3q_t\;
d^3p_t}{(2\pi\hbar)^3}\right)^N\epsilon_0\delta(E-\hat
H_N)}\label{boxM},
\end{equation}
i.e. the volume of ${\cal{M}}$ by that of its closure
$\overline{\cal{M}}$.  In detail we perform the following steps:
\begin{equation}
M_\delta(t,t_0):=<\!G\!>_\delta
*\mbox{vol}_{box,\delta}[{\cal{M}}(t,t_0)],\label{boxM1}
\end{equation}
to obtain $\mbox{vol}_{box,\delta}[{\cal{M}}(t,t_0)]$ we cover the
$d$-dim.\ sub-manifold ${\cal{M}}(t,t_0)$, here with $d=(6N-1)$, of
the phase space by a grid with spacing $\delta$ and count the number
$N_\delta\propto \delta^{-d}$ of boxes of size $\delta^{6N}$, which
contain points of ${\cal {M}}(t,t_0)$. This is illustrated by
fig.(\ref{spaghetti}). Then
$\mbox{vol}_{box,\delta}[{\cal{M}}(t,t_0)]:= \delta^d
N_\delta[{\cal{M}}(t,t_0)]$ and $<\!G\!>_\delta$ is the average of
$\frac{\epsilon_0}{N!(2\pi\hbar)^{3N}||\nabla\hat H||}$ over these
non-empty boxes of size $\delta$.

The \underbar{$\lim$}$_{\delta\to
0}\mbox{vol}_{box,\delta}[{\cal{M}}(t,t_0)]$ is the box-counting
volume of ${\cal{M}}(t,t_0)$ which is the same as the volume of its
closure $\overline{{\cal{M}}(t,t_0)}$, see below:
\begin{eqnarray}
\mbox{vol}_{box}[{\cal{M}}(t,t_0)]&:=&
\underbar{$\lim$}_{\delta\to 0}
\delta^d N_\delta[{\cal{M}}(t,t_0)]\label{boxvol}\\
\lefteqn{\mbox{with }\underbar{$\lim *$}=\inf[\lim *]\mbox{ and write
symbolically:}} \nonumber\\
\lim_{\delta\to 0}M_\delta(t,t_0)&=:&
\displaystyle{B_d\hspace{-0.5 cm}
\int}_{\{q_0\{q_t,p_t\}\in V_1\}}
{\frac{1}{N!}\left(\frac{d^3q_t\;
d^3p_t}{(2\pi\hbar)^3}\right)^N\epsilon_0\delta(E-\hat
H_N)},
\end{eqnarray}
where $\displaystyle{B_d\hspace{-0.5 cm}\int}$ means that this
integral should be evaluated via the box-counting volume (the limit of
expression \ref{boxM1} with the use of \ref{boxvol}) here with
$d=6N-1$. This is illustrated by the following figure.
%%%%%%%%%%%%%%%

The volume of phase space covered by $M_\delta(t,t_0)$ is $\ge
W(E,N,V_1)$. For finite times because of Liouville's theorem
eq.(\ref{liouville}), see also section (\ref{EPSformulation}), we have
\begin{equation}
\underbar{$\lim$}_{\delta\to 0}M_\delta(t,t_0)=W(E,N,t_0,t_0)=W(E,N,V_1)
\end{equation}
At $t\to\infty$ the two limits $\delta\to 0,t\to\infty$ do in general
not commute and as assumed by Gibbs the manifold
${\cal{M}}(t\to\infty)$ becomes dense in the new micro-canonical
manifold ${\cal{E}}(V_2)$.  Then
\begin{equation}
\underbar{$\lim$}_{\delta\to 0}\lim_{t\to\infty}M_\delta(t,t_0)=W(E,N,V_2)
> W(E,N,V_1).
\end{equation}
{\em This is the Second Law of Thermodynamics.} The box-counting is
also used in the definition of the Kolmogorov entropy, the average
rate of entropy gain~\cite{falconer90,crc99}.

The box-counting ``measure'' is analogous to the standard method to
determine the fractal dimension of a set of points~\cite{falconer90}
by the box-counting dimension:
\begin{equation}
\dim_{box}[{\cal{M}}(t,t_0)]:=\underbar{$\lim$}_{\delta\to 0}
\frac{\ln{N_\delta[{\cal{M}}(t,t_0)]}}{-\ln{\delta}}.
\end{equation}

Like the box-counting dimension, the box-counting ``measure'' has the
peculiarity that it is equal to the measure of the smallest {\em
closed} covering set. E.g.: The box-counting volume of the set of
rational numbers $\mbox{vol}_{box}\{{\bf Q}\}$ between $0$ and $1$ is
$=1$, and thus equal to the measure of the {\em real} numbers, c.f.
Falconer~\cite{falconer90} section 3.1.  This is the reason why the
box-counting ``measure'' is not a measure in its mathematical
definition because then we should have
\begin{equation}
\mbox{vol}_{box}\left[\sum_{i\subset\{\bf  Q\}}({\cal{M}}_i)\right]=
\sum_{i\subset\{\bf  Q\}}\mbox{vol}_{box}[{\cal{M}}_i]=0,
\end{equation}
therefore the quotation marks for the box-counting ``measure'', c.f.\
appendix~\ref{app}.

Coming back to the the end of section (\ref{EPSformulation}), the
volume $W(A,B,\cdots,t)$ of the relevant ensemble, the {\em closure}
$\overline{{\cal{M}}(t,t_0)}$ must be ``measured'' by something like
the box-counting ``measure'' (\ref{boxM}) with the box-counting
integral $\displaystyle{ B_d\hspace{-0.5 cm}\int}$, which must
replace the integral in eq.(\ref{phasespintegr}). Due to the fact
that the box-counting volume is equal to the volume of the smallest
closed covering set, the new, extended, definition of the phase-space
integral eq.(\ref{boxM}) is for compact sets like the equilibrium
distribution ${\cal{E}}$ identical to the old one
eq.(\ref{phasespintegr}) and our redefinition of the phase-space
integral by box-counting changes nothing for equilibrium statistics.
Therefore, one can simply replace the old Boltzmann-definition of the
number of complexions and with it of the entropy by the new one
(\ref{boxM}) of course with the understanding that {\em the closure
operation with the box-counting volume (\ref{boxM}) should be done
after the times were specified.}
\section{Conclusion}\label{conclus}
In this paper I showed that Boltzmann's principle
eq.(\ref{boltzmentr1}) covers in a simple and straight way both of
Lebowitz's central issues of statistical
mechanics~\cite{lebowitz99a}. Earlier formulations of these ideas can
be found in \cite{gross178,gross180}. Lebowitz emphasises the
neccessity of self-averaging for thermodynamics which describes the
{\em typical} outcome of a macroscopic measurement. This can only be
expected for large systems, in the thermodynamic limit. However, {\em
there are many situations where even large systems are not
sef-averaging.} E.g. at phase transitions of first order. Moreover, a
whole world of non-extensive systems, like the ``Small'' systems,
show broad, often not single peaked, phase-space distributions. E.g.
in scattering experiments on nuclei or atomic clusters an average
over millions of events is taken. Thus the whole distribution in the
accessible phase space is measured.  These are certainly the most
interesting situations. An {\em extension} of statistical mechanics
to cover also these is demanded.

Macroscopic measurements $\hat{M}$ determine only a very few of all
$6N$ $dof$.  Any macroscopic theory like thermodynamics deals with
the {\em area} $M$ of the corresponding closed sub-manifolds
$\overline{\cal{M}}$ in the $6N$-dim.\ phase space not with single
points.  The averaging over ensembles or finite sub-manifolds in
phase space becomes especially important for the micro-canonical
ensemble of a {\em finite} or any other not self-averaging system.

Because of this necessarily coarsed information, macroscopic
measurements, and with it also macroscopic theories are unable to
distinguish fractal sets ${\cal{M}}$ from their closures
$\overline{\cal{M}}$. Therefore, I make the conjecture: the proper
manifolds determined by a macroscopic theory like thermodynamics are
the closed $\overline{\cal{M}}$. However, an initially closed subset
of points at time $t_0$ does not necessarily evolve again into a
closed subset at $t=\infty$ {\em and the closure operation must be
  explicitely done after setting the times in order to obtain a
quantity that is relevant for a macroscopic theory and can be
compared to thermodynamics}. As the closure operation and the
$t\to\infty$ limit do not commute, the macroscopic dynamics becomes
irreversible.

Here is the origin of the misunderstanding by the famous reversibility
paradoxes which were invented by Loschmidt~\cite{loschmidt76} and
Zermelo~\cite{zermelo96,zermelo97} and which bothered Boltzmann so
much~\cite{cohen97,cohen00}. These paradoxes address to trajectories
of {\em single points} in the $N$-body phase space which must return
after Poincar\'{e}'s recurrence time or which must run backwards if all
momenta are exactly reversed.  Therefore, Loschmidt and Zermelo
concluded that the entropy should decrease as well as it was
increasing before. The specification of a single point in $6N$-dim
phase-space and the reversion of all its $3N$ momentum components
demands of course a {\em microscopic exact} specification of all $6N$
degrees of freedom not a determination of a few macroscopic degrees of
freedom only. No entropy is defined for a single point.
Thermodynamics is addressed to the whole manifold, ensemble of systems
with the same macroscopic constraints. The ensemble develops
irreversibly even though the underlining Newtonian dynamics of each
phase-space point is fully reversible. It is highly unlikely that {\em
  all} points in the ensemble ${\cal{M}}(t,t_0)$ have commensurable
recurrence times so that they can return {\em simultaneously} to their
initial positions.  {\em Once the manifold has spread over the larger
  phase space it will never return.}

Also other misinterpretation of Statistical Mechanics are pointed out:
The existence of phase transitions and critical phenomena are {\em
  not} linked to the thermodynamic limit. They exist clearly and
sharply in ``Small'', non-extensive systems as well. Closed
non-extensive Hamiltonian systems at equilibrium do {\em not} follow
Tsallis non-extensive statistics~\cite{tsallis88}. Boltzmann's
principle describes the equilibrium and the approach of the
equilibrium of extensive as well of non-extensive Hamiltonian systems.

By our derivation of micro-canonical Statistical Mechanics for finite,
eventually ``Small'' systems various non-trivial limiting processes
are avoided.  Neither does one invoke the thermodynamic limit of a
homogeneous system with infinitely many particles nor does one rely on
the ergodic hypothesis of the equivalence of (very long) time averages
and ensemble averages. As Bricmont~\cite{bricmont00} remarked
Boltzmann's principle is the most conservative way to Thermodynamics
but more than that it is the most straight one also.  {\em The single
  axiomatic assumption of Boltzmann's principle, which has a simple
  geometric interpretation, leads to the full spectrum of equilibrium
  thermodynamics including all kinds of phase transitions and
  including the Second Law of Thermodynamics.}

In this paper, I take the fact serious that Thermodynamics as well as
any other {\em macroscopic} theory handles ensembles or sub-manyfolds
and {\em not} single points in phase-space.  {\em Thus the use of
  ensemble averages is justified directly by the very nature of
  macroscopic (incomplete) measurements}. Entropy $s(e,n)$ is the
natural measure of the geometric size of the ensemble. With the
Boltzmann definition of $s(e,n)$ Statistical Mechanics becomes a {\em
  geometric} theory. The topology of its curvature indicates all
phenomena of phase transitions independently of whether the system is
``Small'' or large.

Coarse-graining appears as natural consequence of the
ensemble-nature. The box-counting method mirrors the averaging over
the overwhelming number of non-determined degrees of freedom. Of
course, a fully consistent theory must use this averaging explicitly.
Then one would not depend on the order of the limits $\lim_{\delta\to
0}\lim_{t\to\infty}$ as it was assumed here. Presumably, the rise of
the entropy can then already be seen at finite times when the
fractality of the distribution in phase space is not yet fully
developed. The coarse-graining is no more a mathematical ad hoc
assumption. It is the necessary consequence of the averaging over the
$6N-M$ uncontrolled degrees of freedom. Moreover the Second Law in
the EPS-formulation of Statistical Mechanics is not linked to the
thermodynamic limit as was thought up to now
~\cite{lebowitz99a,lebowitz99}.

In this paper I did not contribute anything to the problem of
describing irreversible thermodynamics of stationary dissipative
systems as it is discussed e.g.\ by Gilbert and
Dorfman~\cite{gilbert99,gilbert00}, Rondoni and
Cohen~\cite{rondoni00}.  As mentioned already dissipation does not
exist in the microscopic dynamics. It is not clear to me how far the
inclusion of dissipation predefines the arrow of time already which
should have been deduced from the theory.  The main problem for me
was the derivation of irreversibility from fully time reversible
microscopic dynamics under maximally clear conditions, i.e. of a
microcanonical closed system.  

Gaspard~\cite{gaspard97,gaspard98} considers systems obeying a
dynamics that preserves the phase-space volume, i.e satisfying
Liouville's theorem, but under non-equilibrium steady state
conditions. Similarly to the present approach he had to coarse grain
(width $\delta$) the accessible phase space.  In conformity to the
standard view of thermodynamics being based on the thermodynamic
limit~\cite{lebowitz99} he then proves the rise of the entropy {\em
  after the limits} (in that order) $\delta\to 0,V\to \infty$.
However, in this limit also the Poincar\'{e} recurrence time becomes
infinite and Zermelo's piercing argument becomes blunted.  So in this
approach Gaspard cannot treat our problem of the Second Law in a {\em
  finite closed} Hamiltonian system which seems to me to be the heart
of the reversibility paradox.

\section{Acknowledgment}
Thanks to A.Ecker for mathematical advises and to J.Barr\'{e},
J.E.Votyakov and V.Latora who had useful suggestions to improve the
text. I am grateful to S.Abe, M.Baranger, E.G.D.Cohen, P.Gaspard,
R.Klages, A.K.  Rajagopal and M.Pettini for many illuminating
discussions.
\section{Appendix}\label{app}
In the mathematical theory of fractals~\cite{falconer90} one usually
uses the Hausdorff measure or the Hausdorff dimension of the
fractal~\cite{crc99}.  This, however, would be wrong in Statistical
Mechanics.  Here I want to point out the difference between the
box-counting ``measure'' and the proper Hausdorff measure of a
manifold of points in phase space. Without going into too much
mathematical details I can make this clear again with the same example
as above: The Hausdorff measure of the rational numbers $\in[0,1]$ is
$0$, whereas the Hausdorff measure of the real numbers $\in[0,1]$ is
$1$. Therefore, the Hausdorff measure of a set is a proper measure.
The Hausdorff measure of the fractal distribution in phase space
${\cal{M}}(t\to\infty,t_0)$ is the same as that of ${\cal{M}}(t_0)$,
$W(E,N,V_1)$. Measured by the Hausdorff measure the phase space volume
of the fractal distribution ${\cal{M}}(t\to\infty,t_0)$ is conserved
and Liouville's theorem applies.  This would demand that
thermodynamics could distinguish between any point inside the fractal
from any point outside of it independently how close it is.  This,
however, is impossible for any macroscopic theory which has only
macroscopic information where all unobserved degrees of freedom are
averaged over.  That is the deep reason why the box-counting
``measure'' must be taken and is a further origin for irreversibility.
%%%\bibliographystyle{unsrt}%{alpha}%{plain} %{unsrt}
%%%\bibliography{gross,othbiba,othbibb,othbibcd,othbibe,othbibf,othbibg,othbibh,othbibij,othbibk,othbibl,othbibm,othbibn,othbibo,othbibp,othbibr,othbibs,othbibt,othbibuw,othbibxz}
%%%\end{document}

\end{document}